\documentclass[12pt, aip,cha,graphicx,amsmath,amssymb,reprint,onecolumn]{revtex4-1}
\usepackage{amssymb, graphics, epsfig, color, ulem, float,graphicx}
\usepackage{amsmath,amstext,bm,verbatim,bbold}
\usepackage{graphicx,subfigure}
\usepackage{epstopdf}
\usepackage[amssymb]{SIunits}
\usepackage{svg}

\usepackage{todonotes}

\newcommand{\emphh}[1]{}

\newcommand{\koffo}{k_{off,1}}

\newcommand{\kofft}{k_{off,2}}

\begin{document}

\title{On the role of mechanosensitive binding dynamics in the pattern formation of active surfaces}
\author{M. Bonati}
\affiliation{Cluster of Excellence Physics of Life, Technische Universit\"at Dresden, Dresden, Germany}
\author{L.D. Wittwer}
\affiliation{Faculty of Mathematics and Informatics, Technische Universtit\"at Bergakademie Freiberg, Freiberg, Germany}
\affiliation{Faculty of Informatics/Mathematics, Hochschule f\"ur Technik und Wirtschaft, Dresden, Germany}
\author{S. Aland}
\email{Corresponding author: sebastian.aland@math.tu-freiberg.de}
\affiliation{Faculty of Mathematics and Informatics, Technische Universtit\"at Bergakademie Freiberg, Freiberg, Germany}
\affiliation{Faculty of Informatics/Mathematics, Hochschule f\"ur Technik und Wirtschaft, Dresden, Germany}
\author{E. Fischer-Friedrich}
\email{Corresponding author: elisabeth.fischer-friedrich@tu-dresden.de}
\affiliation{Cluster of Excellence Physics of Life, Technische Universit\"at Dresden, Dresden, Germany}
\affiliation{Biotechnology Center, Technische Universit\"at Dresden, Dresden, Germany}
\affiliation{Faculty of Physics, Technische Universit\"at Dresden, Dresden, Germany}

\begin{abstract}{
The actin cortex of an animal cell is a thin polymeric layer attached to the inner side of the plasma membrane. It plays a key role in shape regulation and pattern formation on the cellular and tissue scale and, in particular, generates the contractile ring during cell division.
Experimental studies showed that the cortex is fluid-like but highly viscous on long time scales with a mechanics that is sensitively regulated by active and passive cross-linker molecules that tune active stress and shear viscosity. Here, we use an established minimal model of active surface dynamics of the cell cortex 
supplemented with the experimentally motivated feature of mechanosensitivity in cross-linker binding dynamics.
Performing linear stability analysis and computer simulations, we show that cross-linker mechanosensitivity significantly enhances the versatility of pattern formation and enables self-organized formation of stable contractile rings.
}
\end{abstract}
\maketitle
\section{Introduction}
The actin cortex is a thin active polymeric layer that is formed right underneath the plasma membrane of animal cells \cite{salb12, chug18, koen18}. 
It is tethered to the plasma membrane and serves as a mechanical shield as well as an active regulator of cell and tissue shape \cite{salb12, chug18, koen18}. In particular, the actin cortex was shown to play an important role in cell division, cell migration and embryogenesis (e.g. during epithelial folding) \cite{salb12, chug18, koen18}. Therefore, force generation and pattern formation in the actin cortex are an important prerequisite for the successful accomplishment of cellular function in our body. Previous theoretical and experimental studies suggest that corresponding biological pattern formation and shape dynamics of cortical surfaces could, at least in part, be driven through self-organized dynamic processes of cortical constituents, see e.g. \cite{miet19a, miet19b, bois11, kuma14, salb17, wagn16, salb09, bert14, gros19,reym16}.\\
%
The actin cortex is scaffolded by a network of actin polymers decorated by a multitude of associated proteins \cite{salb12, chug18}. In particular, active motor proteins, such as myosin II, cross-link actin polymers and exert tensile forces on them \cite{salb12, chug18}. In this way, motor proteins act as active force dipoles generating active contractile prestress in the cortical layer. 
Also (passive) actin cross-linker proteins such as $\alpha$-actinin were shown to contribute to the generation of active cortical prestress \cite{enno16, fisc16, alva13, koen18, toyo17}. 
Furthermore, the degree of cross-linking of actin networks  was shown to sensitively tune the network's mechanical resistance towards shear deformations \cite{gard04, gard06a}. 
%

In recent years, several experimental studies have provided direct or indirect evidence that actin cross-linkers exhibit mechanosensitive binding dynamics \cite{yao13, schi16, luo13, hoss20, mull20, guo06}. In particular, two scenarios have been suggested for cross-linker binding to actin \cite{hoss20, mull20}; i) tension-induced increase of bond lifetimes - also called catch bond behavior and ii) tension-induced decrease of bond lifetimes - also called slip bond behavior. Therefore, in a  cortex model that captures mechanosensitivity of stress-regulating cortical molecules, the unbinding rate of such molecules needs to depend on the local mechanical stress in the cortical surface. This relationship introduces an additional mechanochemical feedback into the system.

Cortical constituents were observed to undergo dynamic turnover on time scales of seconds to tens of seconds \cite{salb12, chug18, koen18}. In accordance, with that observation, cortical mechanics was shown to be time scale dependent and to be dominantly fluid-like on time scales larger than minutes \cite{fisc16, kelk20}. Accordingly, self-organization and pattern formation in the actin cortex has been modeled as the dynamics of an active viscous fluid, see e.g. \cite{miet19a, miet19b, bert14, salb09, salb17, kuma14, bois11}. Experimental results suggest that the two-dimensional viscosity of the cortical fluid layer is on the order of $10^4-10^5$~$\mu$m$\cdot$Pa~s on long time scales (estimated as the product of parameters  $K_h$ and $\tau_{max}$ from references \cite{fisc16, hoss21}). By comparison, viscosities of the cytoplasm have been estimated in the range of $10^{-2}-10\,$Pa~s{ } \cite{kalw11,dani06}. The ratio between cortical and cytoplasmic viscosity is therefore expected to be on the order of $10^3\,\mu$m or larger. Hence, the hydrodynamic length scale $L_h$ of the coupling between the cytoplasmic bulk and the cortex is expected to be orders of magnitudes larger than typical diameters of animal cells. 

In previous research, Mietke {\it et al.} suggested a minimal model of active spherical cortex surfaces \cite{miet19b}. There, the authors reported the ability of the active model cortex to spontaneously polarize and form patterns. However, spontaneous formation of contractile ring conformations were only reported for the case of small hydrodynamic length scale $L_h=\eta_b/\eta$, where $\eta_b$ is the surface bulk viscosity of the active surface and $\eta$ is the viscosity of the enclosed passive bulk fluid. 
Here, we reconsider pattern formation of this minimal model of active spherical fluid surfaces for the case of large hydrodynamic length scale $L_h$ taking mechanosensitive binding of cortical molecules into consideration as a new feature.

\section{Constitutive equations for an active gel on a curved surface with mechanosensitive molecular regulators}
In the following, we will write down the hydrodynamic equations that govern surface deformation and material flow for a closed active fluid surface in a regime of overdamped dynamics as previously introduced in \cite{miet19b}. The two-dimensional domain of the active surface  is denoted as $\Gamma$.
Let $(s^1, s^2)$ be local coordinates of the manifold $\Gamma$. Correspondingly, $\Gamma$ can be locally parameterized as surface position vectors ${\bf X}(s^1, s^2)\in \mathbb{R}^3$.
An induced basis of the tangent space is given by the vectors ${\bf e}_i=\partial_{i} {\bf X},\, i=1,2$. The surface normal $\bf n$ is  ${\bf n}= {\bf e}_1 \times {\bf e}_2/|{\bf e}_1 \times {\bf e}_2|$.
The tangential velocity of a surface element in $\Gamma$ can then be written as $\textbf{v}_\parallel=v^i {\bf e}_i$.
The normal surface velocity will be denoted as $\textbf{v}_\perp= v_n {\bf n}$. The curvature and metric tensor are defined as $C_{ij}= -{\bf n}\cdot\partial_i {\bf e}_j$ and $g_{ij}={\bf e}_i \cdot {\bf e}_j$, respectively. 
The strain rate tensor of the active surface is 
$v_{ij} = \frac{1}{2}(\nabla_i v_j + \nabla_j v_i) + C_{ij}v_n $ and the in-plane stress tensor reads
\begin{equation}
t_{ij} = 2\eta_s  \,\left(v_{ij}-\frac{1}{2} v_{\,k}^k \, g_{ij}\right) + \eta_b \,\, \, v_{\,k}^k\, g_{ij} + \xi f(c) g_{ij} \quad,
\label{eq:SurfStress}
\end{equation}
where $\eta_s$ and $\eta_b$ are the area shear viscosity and area bulk viscosity of the active fluid and $\xi f(c) g_{ij}$ is an active stress contribution that is tuned by the contractility parameter $\xi$ and a function $f(c)$ that grows monotonically as a function of the local concentration $c$ of molecular regulators. 
We have assumed that surface bending makes negligible in-plane stress contributions using the established membrane approximation of thin closed shells \cite{land86}. 
Furthermore, we did not take into consideration contributions of curvature to active in-plane surface stress. This choice was motivated by experimental findings showing that cell surface tension is largely independent of the degree of cell squishing in a  cell confinement assay in spite of increasing curvature of the unsupported cell surface \cite{fisc14}. \\
The force balance equations in tangential and normal direction are given by
\begin{eqnarray}
\nabla_i t^{i}_j &=& -f^{ext}_j, \qquad
C^{ij}t_{ij}=f^{ext}_n  \quad.   \label{eq:ForceBal}
\end{eqnarray}
External forces in the above Eq.~\eqref{eq:ForceBal} are assumed to stem from stresses in the inner bulk fluid occupying the space enclosed by the active surface $\Gamma$. Therefore, 
${\bf f}^{ext}=-{\bf n} \cdot \boldsymbol{\sigma}|_\Gamma$, where $\boldsymbol{\sigma}=\eta \left(\nabla {\bf u}+(\nabla {\bf u})^T\right)-p \mathbb{1}$ is the stress tensor, $\eta$ is the viscosity and $\bf u$ is the velocity field in  the bulk fluid. 
The Stokes Equation demands 
\begin{eqnarray}
\eta \Delta {\bf u} =\nabla p\quad . \label{eq:StokesEq}
\end{eqnarray}
Furthermore, we impose bulk fluid incompressibility, i.e. $\nabla\cdot {\bf u}=0$ and no slip boundary conditions at the interface with the active surface, i.e. ${\bf u}_{|\Gamma} = {\bf v}_\perp + {\bf v}_\parallel$.

The surface concentration field $c$ of the molecular regulator species is anticipated to undergo a dynamic evolution due to binding and unbinding events that give rise to an exchange with a pool of free molecules in the inner (cytoplasmic) bulk volume $\Omega$. We anticipate that this inner molecular pool is well-stirred due to fast bulk diffusion. By imposing particle conservation, the concentration in the inner bulk is therefore given by $c_{free}=\left(N_{tot}-\int_\Gamma c\, d\Gamma\right)/\int_\Omega d\Omega$, where $N_{tot}$ is the overall number of molecules in the system. 
The time evolution of the concentration field $c$ is modeled as 
\begin{eqnarray}
 \partial_t c + \nabla_i(c v^i) + C^k_{\,k}v_n c - D\Delta_\Gamma c &=& k_{on}c_{free}- k_{off}(t^i_{\,i})c \quad,
 \label{eq:ConcChange}
\end{eqnarray}
where $D$ is the diffusion constant in the active surface, $k_{on}$ is a binding rate of free molecules and $k_{off}(t^i_{\,i})$ is an unbinding rate of surface-associated molecules. 
Furthermore, the term $\nabla_i(c v^i)$  accounts for concentration changes due to tangential advective flux on the surface and the term $C^k_{\,k}v_n c$ captures concentration changes by dilution/condensation through normal surface movement in the presence of curvature. 
Motivated by experimental observations, see e.g. \cite{schi16, hoss20, mull20, luo13}, we anticipate that the unbinding rate of molecular regulators of active stress depends on the local trace of the in-plane stress $t^i_{\,i}$. In particular, we make the choice of a Bell model dependence 
\begin{equation}
k_{off}= k_{off}^0 \exp(-\left(t^i_{\,i}-\sigma_0\right)/\sigma_c)
\label{eq:BellModel}
\end{equation}
where $\sigma_0=2\xi f(c_0)$ is the trace of the stress in the steady state ($c=c_0$) and $\sigma_c$ is a characteristic stress that characterizes the nature of the load dependence of the molecular regulator.  A decrease of the unbinding rate with stress, i.e. $\sigma_c>0$, is known as catch bond behavior while the increase of $k_{off}$ in dependence of stress, i.e. $\sigma_c<0$, is known as slip bond behavior \cite{thom08, mars03}.\\
In the case of the presence of more than one molecular regulator species, the above description of active hydrodynamics can be generalized in a straightforward manner using an active stress contribution $\xi f({\bf c})$, where $f({\bf c})$ is a function of a vector of concentration fields with one vector component for each molecular regulator. Furthermore, time evolution equations in the form of Eq.~\eqref{eq:ConcChange} must be defined for the concentration fields of all molecular regulator species.

\section{Pattern formation of an active spherical surface with mechanosensitive  molecular regulators}
In the following, we will discuss self-organized pattern formation and surface deformation for the active hydrodynamics as described by Eqn.~\eqref{eq:ForceBal} and \eqref{eq:ConcChange} emerging from a spatially uniform steady state of a spherical shape with radius $R_0$. Therefore, initially, the uniform surface concentration is $c_0=3N_{tot}k_{on}/(4\pi R_0^2)\times 1/(3k_{on}+R_0 k_{off}^0)$ and velocities vanish on the surface and in the bulk. \\
To identify the stability of this steady state, we  perturb the concentration field and the radius of the active surface. As spherical harmonics 
$Y_{\ell m}(\theta, \varphi)\, (\ell=0,1, \ldots, \infty, m=-\ell, \ldots, \ell)$  
are an orthonormal basis of the function space on the sphere, we can write 
$\delta c = \sum_{\ell,m} \delta c_{\ell m}Y_{\ell m}(\theta, \varphi)$ and 
$\delta R= \sum_{\ell,m} \delta R_{\ell m}Y_{\ell m}(\theta, \varphi)$
where $\theta$ and $\varphi$ denote the polar and azimuthal angle of the sphere, respectively. 
Correspondingly, we expand the emerging tangential velocity field in the form
$
\delta \textbf{v}_\parallel = \sum_{\ell,m} ( \delta v_{\ell m}^{(1)}\mathbf{\Psi}^{(lm)}(\theta, \varphi) + \delta v_{\ell m}^{(2)}\mathbf{\Phi}^{(lm)}(\theta, \varphi) ),
$
where $\mathbf{\Psi}^{(\ell m)}(\theta, \varphi)$ and $\mathbf{\Phi}^{(\ell m)}(\theta, \varphi)$ are the tangential vector spherical harmonic functions. 
Eqn.~\eqref{eq:ForceBal}, \eqref{eq:StokesEq} and \eqref{eq:ConcChange} are then expanded in the perturbations $\delta c, \delta R$ and $\delta \textbf{v}_\parallel$ to linear order and the growth rates $\lambda_{\ell}$ of each harmonic mode are determined as previously described by Mietke {\it et al.} \cite{miet19b} (see Supplementary Section~1). In the linear regime, mode amplitudes $\delta c_{\ell m}, \delta R_{\ell m}$ and $\delta v_{\ell m}$ change in time proportional to $\exp(\lambda_{\ell}t)$, i.e. if $\lambda_{\ell }>0$, the perturbation mode grows and if $\lambda_{\ell }<0$ the perturbation mode shrinks over time. (Note that the growth rate is independent of mode parameter $m${ } \cite{miet19b}.) For the limit case of vanishing bulk viscosity $\eta$ (i.e. diverging hydrodynamic length scale $L_h=\eta_b/\eta$), an analytical expression for the growth rate is given in Supplementary Section~2. 
Furthermore, in this limit case, we find that the induced change of the trace of the stress tensor $\delta t^i_{\, i}$  is to first order
\begin{eqnarray}
\delta t^i_{\, i}&=&2\eta_b v^i_{\, i}+2 \xi  f'({ c}_0) \delta  c   \notag\\
&=& \frac{2 \left(\ell^2+\ell-2\right) \xi ^2 f\left({ c}_0\right)  \eta_s \big( f'\left({ c}_0\big) \delta { c}_{\ell m} Y_{\ell m}(\theta ,\varphi )\right)}{\xi  f\left({ c}_0\right) \left(\ell (\ell+1) \eta_b+\left(\ell^2+\ell-2\right) \eta_s\right)+4 \lambda_\ell  \eta_b \eta_s} \quad,
\label{eq:StressTracePert}
\end{eqnarray}
for harmonic perturbations with mode degree $\ell$.
Here, $\lambda_\ell$ is the corresponding growth rate. 
Accordingly, the trace of the stress tensor does not change through the dynamics for modes with $\ell=1$ (up to linear order). This behavior results from the lack of area shear contributions for $\ell=1$ perturbations for vanishing bulk viscosity (see Supplementary Section~3).
We conclude that the unbinding rate $k_{off}(t^i_{\, i})$ stays constant for these modes (in space and time) and the associated mechanosensitivity of the molecular regulator does not affect the binding dynamics (up to linear order).
Correspondingly, the stress-amplified binding as mediated by a catch-bond mechanosensitivity of a molecular regulator can be used to specifically amplify modes with $\ell>1$. 

\subsection{One mechanosensitive regulator species}
\begin{figure}[ht]
\includegraphics[width=14cm]{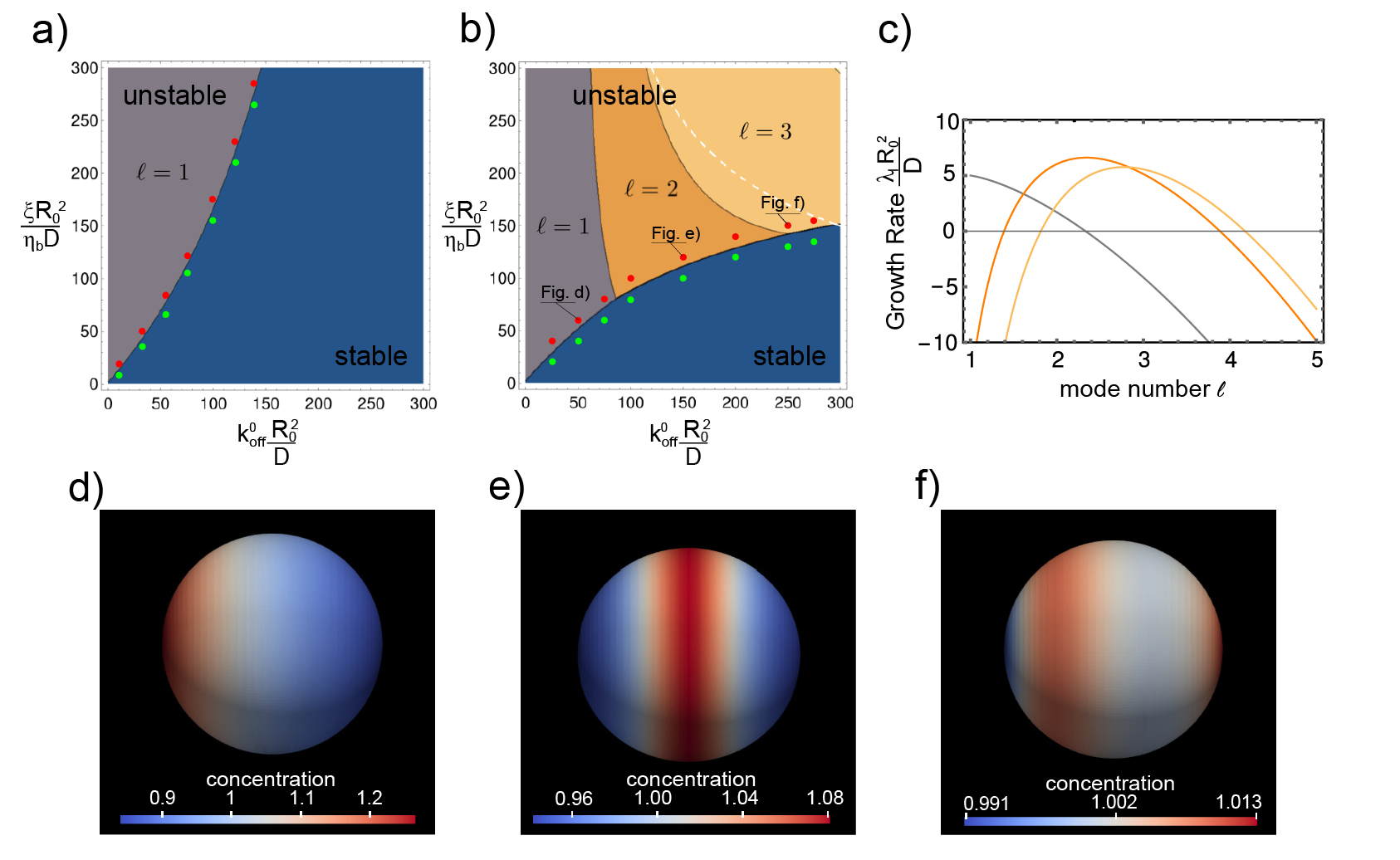}
\caption{\textbf{
Linear stability analysis of self-organized pattern formation of an active spherical surface  with one mechanosensitive molecular regulator according to model Eqn.~\eqref{eq:ForceBal}-\eqref{eq:BellModel}.}
a) Phase diagram of the linear stability of an active spherical surface with a slip bond molecular regulator ($\sigma_c<0$) in dependence of dimensionless unbinding rate $k_{off}^0 R^2_0/D$ and dimensionless active contractility $\xi R_0^2/(\eta_b D)$. 
The uniform steady state is stable (blue region) or unstable with dominant eigenmode  $\ell=1$ (grey region).
b) Phase diagram of the linear stability of an active spherical surface with a catch bond regulator  ($\sigma_c>0$). The uniform steady state is stable (blue region) or unstable with dominant eigenmodes $\ell=1$ (grey), $\ell=2$ (orange) or $\ell=3$ (light orange).
The white dashed line indicates the stability boundary for homogeneous concentration perturbations ($\ell=0$ mode).
This stability boundary can be moved upwards by increasing the chosen value of $k_{on}$. 
a,b) Red and green dots represent simulation results illustrating the consistency of the numerical model. Green dots indicate stable steady states while red dots indicate growing perturbations over time. 
c) Dimensionless growth rates of different pattern modes in dependence of mode degree $\ell$ for three parameter choices corresponding to marked points in panel b (curve colors match colors of corresponding stability regions).
Each curve corresponds to a different dominant mode (grey curve: $\ell=1$, orange curve: $\ell=2$, light orange curve: $\ell=3$). 
d-f) Simulation snapshots of concentration distributions  for parameter choices as indicated in panel b  at simulation time $\hat t=1$. 
Initial (dimensionless) concentrations were chosen as the uniform  steady state concentration $1$ perturbed by an axisymmetric concentration profile with random values in the interval $\left[-5\cdot 10^{-3}, 5\cdot 10^{-3}\right]$ (see Supplementary Section~5). As predicted by the linear stability analysis (see panel b), different pattern modes are chosen to be amplified by the dynamics.
Dimensionless parameters used were 
$k_{on} R_0/D=200$,
$\eta_s/\eta_b=0.25$, $\eta R_0/\eta_b=0.1$,
$\Delta \hat t  = 10^{-5}$
and $\sigma_c R_0^2/(\eta_b D)=-100$ (a) and $\sigma_c R_0^2/(\eta_b D)=100$ (b). 
In addition, for panels d-f, we chose $\xi R_0^2/(\eta_b D)=60, 120,150$ and $k_{off}^0 R_0^2/D=50, 150, 250$, respectively. 
\label{fig:Fig1}
}
\end{figure}

Using linear stability analysis, we first considered the stability of harmonic modes with $\ell\ge 1$ for one molecular regulator species in the limit of large hydrodynamic length scale $L_h$ with the specific choice $f(c)=2c^2/(c_0^2+c^2)$. 
For both, catch and slip bond mechanosensitivity, we find that the steady state becomes unstable if the contractility parameter $\xi$ is sufficiently large  (see Fig.~\ref{fig:Fig1}a,b). For slip bond mechanosensitivity ($\sigma_c<0$), the mode with $\ell=1$ is predicted to be always dominant, i.e. its growth rate is always the largest (see Fig.~\ref{fig:Fig1}a). This behavior is analogous to the situation of no mechanosensitivity \cite{miet19b}. For catch bond mechanosensitivity ($\sigma_c<0$), we find that the versatility of predicted patterns is greatly enriched in that parameter regimes with different dominant modes exist. In particular, parameters can be chosen such that arbitrary $\ell$-modes can become dominant (see Fig.~\ref{fig:Fig1}b,c). A mode is denoted as dominant if the real part of the growth rate is maximal at its corresponding degree $\ell$. 

To examine the behaviour of the system away from the linear stability region, we solved the Eqn.~\eqref{eq:ForceBal},\eqref{eq:StokesEq} and \eqref{eq:ConcChange} numerically in an axisymmetric scheme using a finite-element framework and the Arbitrary-Lagrangian-Eulerian (ALE) method (see Supplementary Section~5) \cite{witt22, Vey07, witkowski2015software}. We find that  numerical simulations agree with predictions of the linear stability analysis with regards to stability of the uniform state  (see green and red dots in Fig.~\ref{fig:Fig1}a,b). 
Furthermore, numerical simulations reproduce the amplification of  different harmonic modes  from random initial perturbations of the steady state as predicted by the linear stability analysis for catch bond molecular regulators (see Fig.~\ref{fig:Fig1}b,d,e,f).

\subsection{Two molecular regulator species with opposite mechanosensitivity}
\begin{figure}[ht]
\centering
\includegraphics[width=14cm]{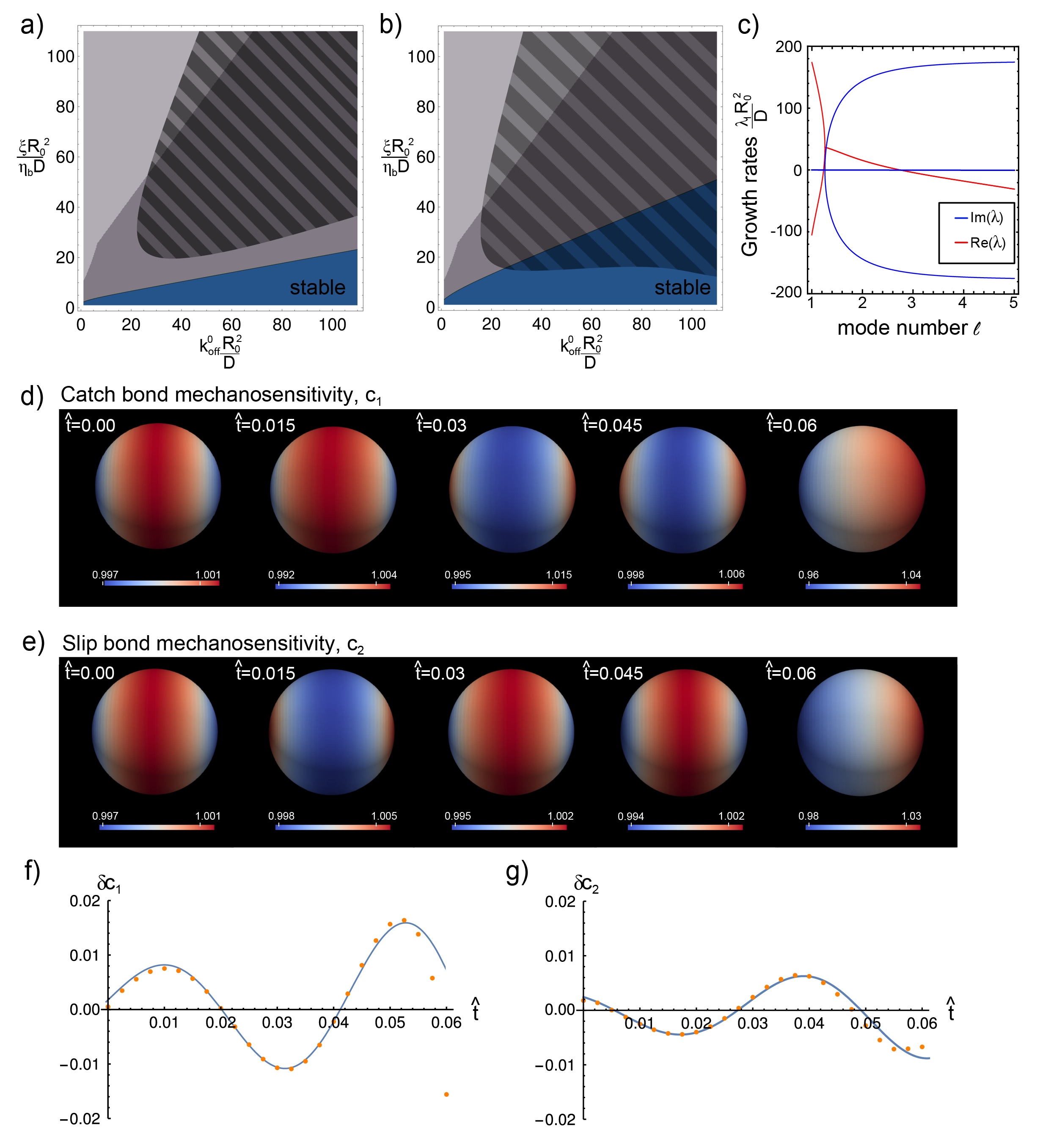}
\caption{
\textbf{ Self-organized pattern formation of an active spherical surface with  catch and slip bond molecular regulators of opposite mechanosensitivity.}
a,b) Phase diagrams of the linear stability of an active spherical surface with catch and slip bond molecular regulators in dependence of dimensionless unbinding rate $k_{off}^0 R^2_0/D$ and dimensionless active contractility $\xi R_0^2/(\eta_b D)$ for two different values of internal viscosity (a: $\eta R_0/\eta_b=0$, b: $\eta R_0/\eta_b=0.1$). Blue region: the uniform steady state is stable, grey regions: the uniform steady state is unstable and the eigenmode with degree $\ell=1$ is dominant. In the light grey area, also the eigenmode with $\ell=2$ is unstable but not dominant. The striped region further indicates a parameter region where the imaginary part of the $\ell=2$ mode growth rate does not vanish. Therefore, in the dashed parameter regime, oscillations of concentration perturbations proportional to the $\ell=2$ mode are predicted.
c) Real (red) and imaginary (blue) part of dimensionless growth rates of different pattern modes in dependence of mode number $\ell$ according to the linear stability analysis. (For one of the eigenvalues, the real part was  below $-300$ for all $\ell$ and is not shown.) 
Parameters are equivalent to parameters of the simulation presented in panel d and e.
d,e) Concentration patterns emerging from a small $\ell=2$ mode perturbation (amplitude $-5\cdot 10^{-3}$). Note that the depicted concentration range grows over time. 
For increasing times, the pattern switches between an aggregation at the center (ring conformation) to aggregations at both poles. 
Once numerical errors build up, there is a spontaneous left-right symmetry breaking and the system is driven towards the dominant $\ell=1$ mode.
f,g) Concentration perturbations at the surface center as predicted by linear stability analysis (blue line) and as obtained from simulations (orange dots). When the system starts to evolve towards the $\ell=1$ mode due to spontaneous symmetry breaking, there is an emerging discrepancy between linear stability analysis and simulations ($\hat t>0.05$).
Parameters used were 
$k_{on,1}R_0/D=300$, $k_{on,2}R_0/D=200$, 
$\sigma_{c,1} R_0^2/(\eta_b D)=125$, $\sigma_{c,2} R_0^2/(\eta_b D)=-12.5$,
$\eta_s/\eta_b=0.25$,
$\eta R_0/\eta_b=0.1$
and $\Delta \hat t = 2.5\cdot 10^{-7}$. 
For simulation results presented in panels d-g, we used in addition $\xi R_0^2/(\eta_b D)=400$, $k_{off,1}^0 R_0^2/D=250$ and $k_{off,2}^0 R_0^2/D=25$.
Here, $D$ is the diffusion constant for both molecular regulators.
\label{fig:Fig2}}
\end{figure}

Some actin cross-linkers such as $\alpha$-actinin-4 and myosin have been shown to exhibit a catch-bond behavior \cite{schi16, hoss20, veig03}. For other actin cross-linkers such as fascin, previous data indicate a slip bond behavior \cite{liel08}. Therefore, it is a plausible scenario that molecular regulators of cortical activity may exhibit both kinds of mechanosensitivity.
In fact, admitting for at least two molecular regulator species  with equal but opposite mechanosensitivity, the richness of pattern forming behaviors can be further extended;  in particular, it is possible to find parameter regimes, where growth rates exhibit a positive real-valued part and a non-vanishing imaginary part indicating oscillatory behavior. \\
In the following, we will consider a concentration field $c_1$ describing a catch bond molecular regulator ($\sigma_1>0$) in combination with a second concentration field $c_2$ describing a slip bond molecular regulator ($\sigma_2<0$). Making the specific choice of  $f(c_1, c_2)=2 c_1 c_2/(\sqrt{c_{0,1} c_{0,2}}(c_1+c_2))$
and $k^0_{off,1}=10 k^0_{off,2}$, we present a phase diagram as predicted by linear stability analysis in  Fig. \ref{fig:Fig2}a (vanishing bulk viscosity) and \ref{fig:Fig2}b (small bulk viscosity $\eta R_0/\eta_b=0.1$). For $\ell>1$, oscillatory dynamics of mode perturbations can be found (see Fig. \ref{fig:Fig2}d,e,f,g). However, the real part of the growth rate of the $\ell=1$ mode is always greater than the growth rate of an oscillatory higher order mode. In accordance with this, we see stationary solutions with one concentration maximum at the pole emerging at longer times (see Fig. \ref{fig:Fig2}d,e).

It is noteworthy that for flat active surfaces, there are no pattern-forming eigenmodes that avoid shear strains in the linear regime. Correspondingly, persistent oscillatory concentration patterns are predicted by the linear stability analysis (see Supplementary Section~4). Making the choice that $D=D_1=D_2$, $\partial_{c_1}\! f({\bf c}_{0})=\partial_{c_2}\! f({\bf c}_{0})$, $c_{0,1}=c_{0,2}$, $\kofft^\prime(\sigma_0) =- \koffo^\prime(\sigma_0)$ and $k_{off,1}^0>k^0_{off,2}$, we find oscillatory solutions  if 
\begin{equation}
\frac{8\eta_s c_{0,1}\xi \partial_1 f ({\bf c}_0)|k_{off,1}^\prime(\sigma_0)|}{(\eta_s+\eta_b)}> (k_{off,1}^0-k^0_{off,2}) \quad,
\end{equation}
where $\sigma_0=2 \xi f({\bf c}_0)$ is the trace of the stress in steady state (see Supplementary Section~4).
The emergence of oscillations in this parameter regime can be understood as follows: perturbative local enrichments of molecular regulators  will give rise to local contractile advection because of locally increased $f({\bf c})$. In the presence of shear deformation, the induced flow will give rise to a local increase of the trace of the stress tensor $t^i_{\,i}$. In turn, the catch bonding property of the first molecular regulator will at first locally increase $c_1$ and local contractility. After some delay ($k_{off,1}^0>k_{off,2}^0$!), the locally increased $\kofft$ of the slip bonding regulator  kicks in reducing the local $c_2$ and correspondingly also local contractility.  Therefore, concentration bumps are first amplified (due to the fast catch bonding regulator) and then shrink with a delay (due to the slow slip bonding regulator) giving rise to oscillatory concentration changes.
It is noteworthy that this effect relies on the presence of a non-vanishing shear viscosity and shear deformation. Correspondingly, oscillatory eigenvalues do not occur for $\eta_s=0$. 
Furthermore, it is noteworthy that the here reported mechanism of oscillatory pattern formation relies on the well-known mechanism of a fast activator (catch-bonding regulator) and a slow inhibitor (slip-bonding regulator) as was previously in depth characterized for reaction-diffusion systems \cite{murr13, cros93}. 


\section{Ring constriction on active spherical surfaces depends on the ratio between shear and bulk viscosity}
During the division of animal cells, an initially round cell constricts through a contractile ring that forms as part of the actomyosin cortex of the cell \cite{woll16, spir17}. This contractile ring is known to be enriched in molecular contractility regulators (myosin motors) \cite{woll16,spir17}. It has been proposed that  mitotic cell constriction might be driven by self-organized pattern formation of the active actomyosin cortex at the cell periphery, see e.g. \cite{miet19a, miet19b, bois11, kuma14, salb17, wagn16, salb09, bert14, gros19,reym16}. Furthermore, myosin motors have been previously suggested to exhibit catch bond mechanosensitivity in their actin binding \cite{schi16, guo06}. 

\begin{figure}[htbp]
\includegraphics[width=14cm]{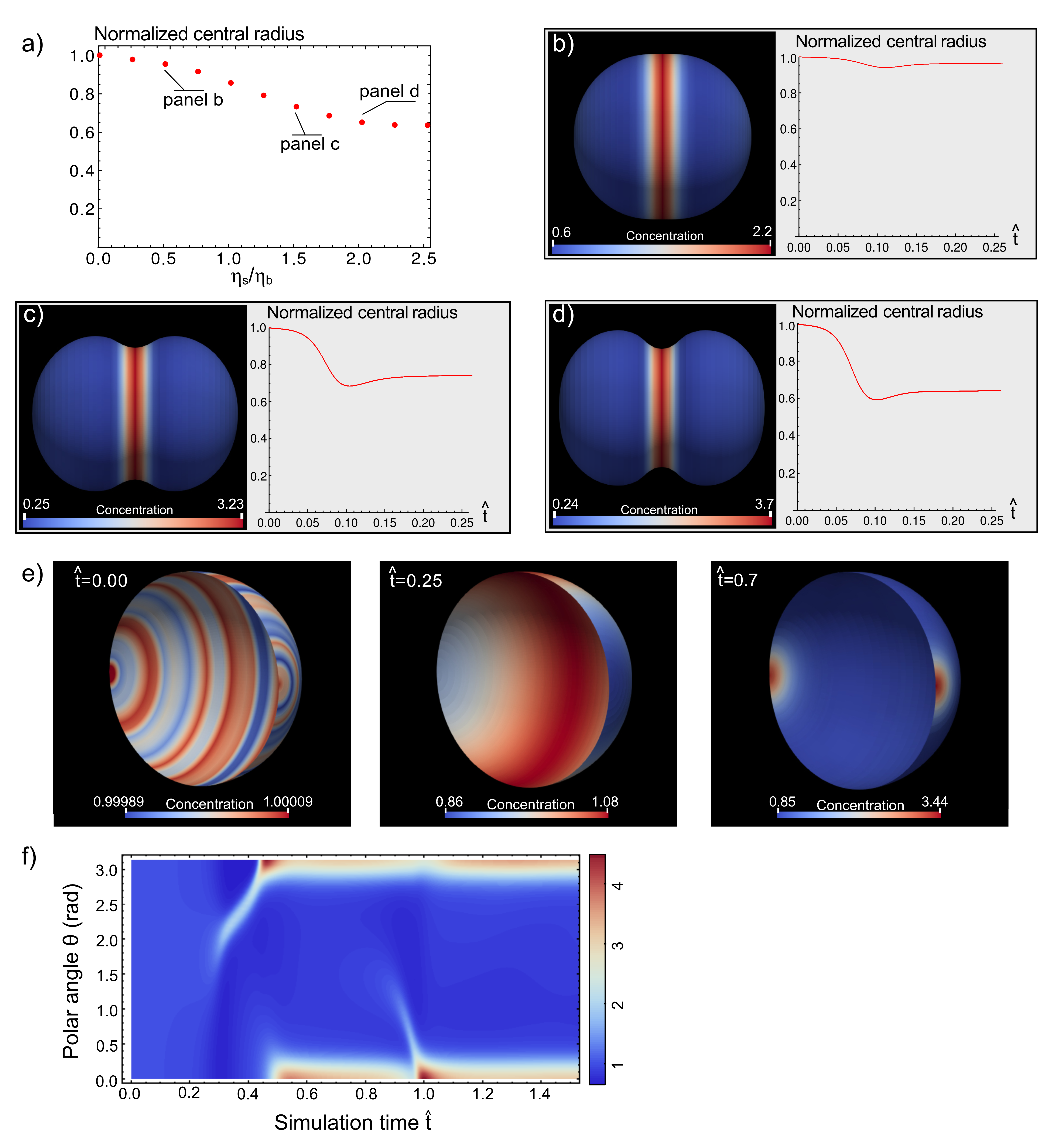}
\caption{
\textbf{The ratio between area shear and bulk viscosity tunes ring constriction.}
a) Steady-state central radius $R(\theta=\pi/2)$  after simulation time $\hat t=0.25$ in dependence of shear viscosity $\eta_s$. Radii are normalized by the initial radius of the sphere $R_0$. Normalized radii below one correspond to constriction of the model cell in the center.
Each simulation was initialized with an $\ell=2$ mode perturbation (amplitude of $-5\cdot 10^{-2}$).
b,c,d) left panels: snapshots of simulations at time point $\hat t=0.25$ corresponding to marked points in panel a. Shear viscosities were chosen as $\eta_s=0.5$ for panel b, $\eta_s=1.5$ for panel c and  $\eta_s=2$ for panel d. 
Right panels: time evolution of the normalized central radius as a function of simulation time.
e) Simulation snapshots for shear viscosity $\eta_s=0.25$ at simulation times ($\hat t=0, 0.33$ and $1$). Here, the initial steady state was perturbed by an axisymmetric concentration profile with random values in the interval $\left[-10^{-4},  10^{-4}\right]$ (leftmost panel, see Supplementary Section~5).
The system is first driven into a ring-shaped concentration accumulation close to the equator (see central panel). Later, the ring slides to the one of the poles (see panel f and Supplemental Movie 1) and, thereafter,  concentration maxima develop at both poles (rightmost panel).
f) Kymograph of concentration along polar angle $\theta$ for the simulation shown in panel e.
Dimensionless parameters used were 
$k_{on}R_0/D=200$, $k_{off}^0 R_0^2/D=100$,
$\eta R_0/\eta_b=0.1$, 
$\sigma_c R_0^2/(\eta_b D)=100$ and $\xi R_0^2/(\eta_b D)=125$  respectively. 
The time step was chosen as  $\Delta \hat t = 2.5\cdot 10^{-6}$.
\label{fig:Fig3}}
\end{figure}

Therefore, we asked to which extent  active spherical surfaces in our model can form a ring constriction by one molecular regulator with catch bond mechanosensitivity.
Overall, in the unstable parameter regime, small ring-shaped concentration perturbations (proportional to a negative $\ell=2$ mode, see e.g. Fig.~\ref{fig:Fig1}e) lead to a surface constriction that is enhanced by the contractility parameter $\xi$, see \cite{witt22}. Here, we show that constriction increases also in dependence of the ratio $\eta_s/\eta_b$ between area shear viscosity and area bulk viscosity (see Fig.~\ref{fig:Fig3}a-d). Intuitively, this can be understood as follows; in a ring-shaped concentration peak, there is a continuous flux from the poles to the ring \cite{witt22}, leading to a unidirectional compression of material in the direction orthogonal to the ring. If shear viscosity is high, shear deformations (i.e. aspect ratio changes of surface elements) tend to be avoided by the dynamics of the system. Correspondingly, unidirectional compression of the material orthogonal to the ring tends to trigger compression along the ring. Therefore, the ring gets smaller in diameter, i.e. it constricts.

For growing values of $\eta_s/\eta_b$, harmonic modes with increasing value of $\ell$ become dominant. In particular, for the parameter choice in Fig.~\ref{fig:Fig3}, mode dominance transitions from the $\ell=2$ mode to the $\ell=3$ mode once $\eta_s/\eta_b >0.66$. 
Hence, we see that large symmetric ring constrictions are not a stationary state but start to evolve to an asymmetric distribution after $\hat t=0.3$ or greater.

We had seen that in certain parameter regimes ring-shaped concentration distributions may emerge from random concentration perturbations of catch bonding cross-linkers (see Fig.~\ref{fig:Fig1}b,e). However, we observed in simulations that, in these cases, the emerging ring is not stable in the nonlinear regime and slips to one of the poles such that, eventually, the system emerges to a steady state conformation with two concentration maxima at the poles (see Fig.~\ref{fig:Fig3}e,f and Movie 1). We conclude, that with the current model of one mechanosensitive catch bond cross-linker in the limit of low cytoplasmic viscosity (large hydrodynamic length scale), we could not find parameters for which we obtained stable constricted rings emerging from random perturbations in a self-organized manner.


\section{Concentration-dependent shear viscosities can enhance the robustness of ring formation}
In the search for mechanisms that would stabilize the `ring' pattern on an active surface, we further investigated modifications of the model that exclusively affected the nonlinear range of concentration and shape perturbations. In particular, we decided to investigate the case of concentration dependent shear viscosities. The concentration dependence of shear viscosity can be motivated by two arguments: 1) It has been observed experimentally that the complex shear modulus increases upon increased cross-linking, see e.g. \cite{gard04, gard06a}. (In particular, cross-linking was associated to a disproportionate increase of shear stiffness over bulk stiffness as can be inferred from the observation of a cross-linking induced  decrease of the Poisson ratio \cite{grea11}.) 2) The second argument is provided by the experimental finding that active stress leads to shear-stiffening of biopolymeric networks as was suggested by {\it in vivo} and  {\it in vitro} studies \cite{fisc16, wang02, koen09, fern06, stam04}. This stress-stiffening might result from the shear-induced nematic alignment of cortical filaments \cite{fisc18, ingb14} or from other stress-induced structural changes \cite{broe11}. (As active stress in the active surface increases monotonically as a function of concentration of molecular regulators, the shear viscosity increase in dependence of active stress can be translated into a viscosity increase in dependence of molecular concentration $c$.)

\begin{figure}[htbp]
\centering
\includegraphics[width=14cm]{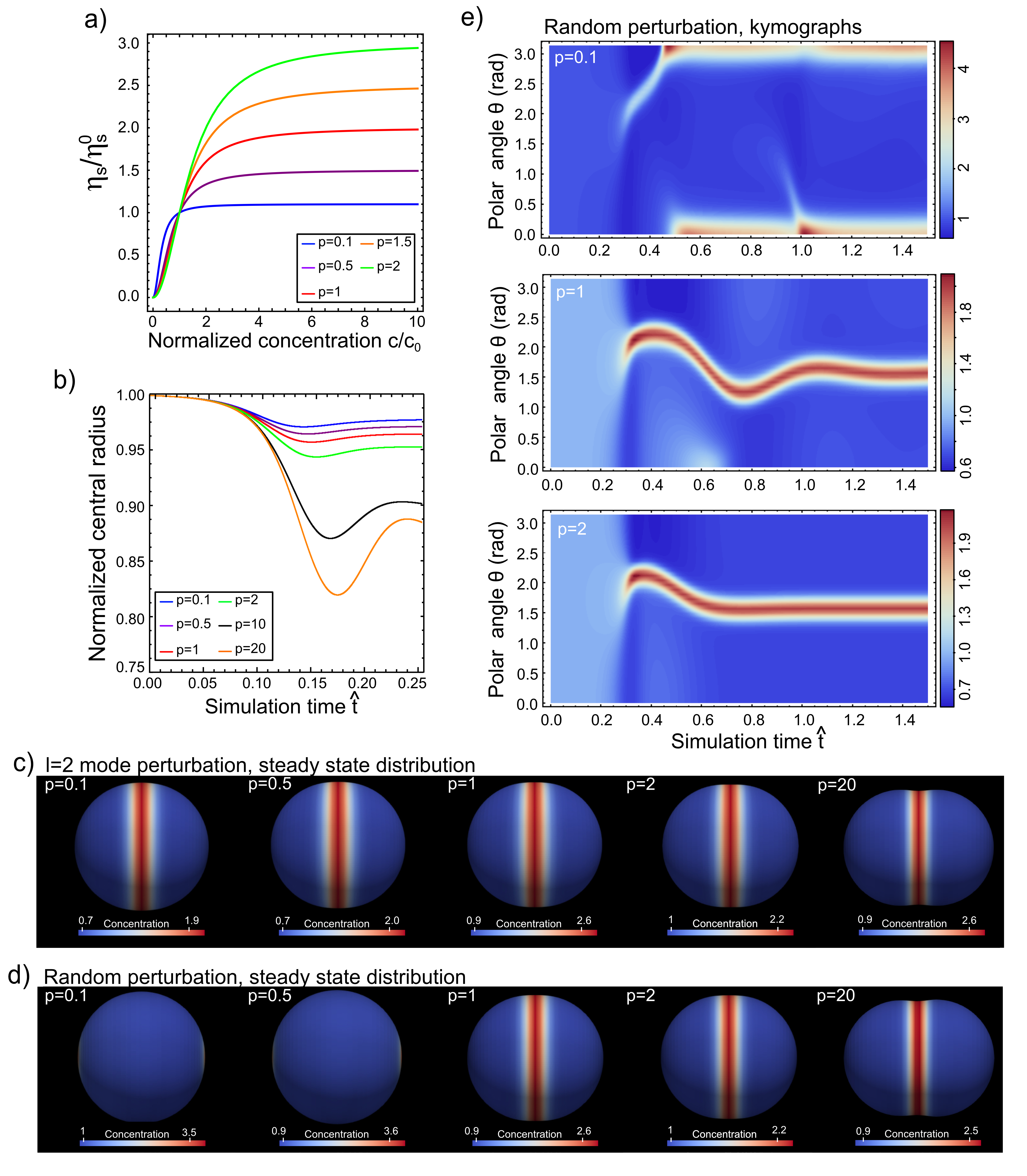}
\caption{
\textbf{Concentration-dependent shear viscosities consolidate ring formation in model cells with catch bonding molecular regulators.}
a) Concentration-dependent shear viscosities $\eta_s=\eta_s^0 (p+1)c^2/(p c_0^2+c^2)$ in dependence of normalized regulator concentration $c/c_0$ with parameter values $p=0.1, 0.5, 1, 1.5$ and $2$ (from bottom to top).
b) Time evolution of the normalized central radius  for different concentration dependencies of shear viscosity according to different parameter choices of $p$ (see panel a). 
Normalized radii below $1$ correspond to central constriction of the model cell.
Each simulation was initialized with an $\ell=2$ mode perturbation (amplitude:  $-5\cdot 10^{-2}$).
c) 
Snapshots of simulation results at time point $\hat t=0.25$ corresponding to curves shown in panel a. Simulations were started with an $l=2$ mode perturbation (shallow ring profile) which is amplified over time to a persistent, ring-shaped concentration accumulation with larger amplitude. %
d) Snapshots of simulation results at time point $\hat t=1.5$ emerging after perturbing the uniform steady state concentration with an axisymmetric concentration profile with random values in the interval $\left[-10^{-4}, 10^{-4}\right]$ (see Supplementary Section~5).
Depending on the level of the concentration dependence of the shear viscosity, the final steady state pattern changes from concentration maxima at the poles (low values of $p$) to a ring-shaped concentration pattern (larger values of $p$).
The full dynamics of the active surface pattern for $p=0.1, 1$ and $2$ is shown in Supplemental Movies 2-4.
e) Kymographs showing the concentration evolution  corresponding to the three simulation snapshots with $p=0.1, 1$ and $2$ in panel d. Additional kymographs showing ring centering in the presence of strong viscosity dependence ($p=10$ and $p=20$) are shown in Fig.~S1 of the Supplement.
Dimensionless parameters used were 
$k_{on}R_0/D=200$, $k_{off}^0 R_0^2/D=100$, 
$\eta_s^0=0.25$, $\eta R_0/\eta_b=0.1$, 
$\sigma_c R_0^2/(\eta_b D)=100$, $\xi R_0^2/(\eta_b D)=125$ and  $\Delta \hat t = 2.5\cdot 10^{-6}$, respectively. 
\label{fig:Fig4}}
\end{figure}

We make the specific choice of $\eta_s=\eta_s^0(p+1) c^2/(p c_0^2+c^2)$, where $p$  is a parameter that tunes the concentration dependence of shear viscosity. Here,  $p=0$ corresponds to the case of no concentration dependence and $p =\infty$ corresponds to the strong concentration dependence $\eta_s=\eta_s^0 c^2/c_0^2$ (see Fig.~\ref{fig:Fig4}a). We find that increasing values of $p$ lead to increasing constriction of ring distributions in our simulations (see Fig.~\ref{fig:Fig4}b,c). 
This is expected from our finding that high ratios of area shear and bulk viscosity $\eta_s/\eta_b$ promote ring constriction, as this viscosity ratio grows with parameter $p$ inside the ring (see Fig.~\ref{fig:Fig3}a and Fig.~\ref{fig:Fig4}a,c).

While for low values of $p$ the contractile ring slips toward the poles at later times (see Fig.~\ref{fig:Fig4}e, top panel and Movie 2), we find that for a sufficiently strong rise of $\eta_s$ with concentration $c$, this ring slippage does not take place anymore (see Fig.~\ref{fig:Fig4}d,e). In fact, we find that initially off-centered ring conformations can be centred over time by the dynamics of the system (Fig.~\ref{fig:Fig4}e, Fig.~S1 and Movie 3 and 4). This behavior can be understood as follows; in the ring, regulator concentrations are high which increases viscosity locally and slows ring dynamics; when the ring is in an off-center position, the neighboring larger segment of the sphere is drained to a lesser extent of molecular regulators through advective flux into the ring. Therefore, in the larger segment, the regulator concentration starts to become higher than in the smaller segment of the sphere on the other side of the ring (see e.g. time point $\hat t = 0.6$ in Fig.~\ref{fig:Fig4}e, mid panel). Therefore, the ring is `fed' more from the side of the larger segment. In turn, the ring moves towards the side of the larger segment over time. If the ring motion is not slow enough, this phenomenon can lead to an overshoot, that means a ring movement beyond the center (see e.g. time point $\hat t = 0.9$ in Fig.~\ref{fig:Fig4}e, mid panel). A damped left-right oscillation of the ring follows until the ring stays in the center (see  Fig.~\ref{fig:Fig4}e, mid panel, and Movie 3).
For sufficiently large values of $p$, viscosities in the ring are so high, that no overshoot happens and the ring centers more quickly (see  Fig.~\ref{fig:Fig4}e, bottom panel, Fig.~S1 and Movie 4).

In spite of the described effect of ring centering, we observed that the emergence of a ring conformation depends on the nature of the random perturbation. Choosing ten different random seeds, we see ring emergence in seven out of ten cases. In the remaining three cases, concentration aggregates start to emerge at the pole or close to the pole. In these cases, two concentration maxima at the pole are forming as steady state conformation. Therefore, we conclude that emergence of a ring conformation from a uniform state requires perturbations that favor ring-shaped concentration patterns over polar concentration peaks.

\section{Discussion}

In this study, we revisited a minimal model of active cortical surfaces of animal cells as introduced by Mietke {\it et al.} \cite{miet19b}. Model cortices with spherical reference geometry were described as closed active viscous surfaces whose mechanics is determined by a shear and bulk surface viscosity $\eta_s$ and $\eta_b$ as well as an active stress contribution adjusted by the local concentration $c$ of molecular regulators  and an activity tuning parameter $\xi$ (see Eq. \eqref{eq:SurfStress}). Furthermore, an internal bulk fluid of significantly lower viscosity $\eta$ was mechanically coupled to the active surface. 
The corresponding hydrodynamic length scale $L_h=\eta_b/\eta$ was chosen to be large as suggested by measurements on cells \cite{fisc16, hoss21}. 

The model by Mietke {\it et al.} was extended by  the new feature of mechanosensitive molecular regulators. In particular, we discussed the case of a molecular unbinding rate $k_{off}$ with an exponential dependence on the trace $t^i_{\,i}$ of the local surface stress tensor (see Eq.~\eqref{eq:SurfStress}). Corresponding exponential growth of the unbinding rate was mimicking a slip bond behavior, while an exponential decline with $t^i_{\,i}$ was mimicking a catch bond behavior of the molecular regulator. Therefore, the lifetime of catch bond regulators at the active surface is increased for higher values of  $t^i_{\,i}$, while the lifetime  of slip bond cross-linkers at the active surface is reduced at increased $t^i_{\,i}$.  

Using linear stability analysis and finite element numerical simulations, we analysed the pattern formation of our model in different parameters regimes.  For the case of one molecular species with slip-bond mechanosensitivity, we find that the harmonic mode with $\ell=1$ always becomes unstable first analogous to the case of stress-independent $k_{off}$ (see Fig.~\ref{fig:Fig1}a) \cite{miet19b}.
By contrast, for the case of one molecular regulator with catch-bond mechanosensitivity,  the spectrum of emerging patterns significantly widens; for arbitrary (spherical) harmonic modes, we found parameter regions in which associated  concentration perturbations are dominantly amplified (Fig.~\ref{fig:Fig1}b). 
This effect originates from the growth rates of higher order modes ($\ell>1$) being increased through the stress-dependence of $k_{off}$ in the catch bond case. On the other hand, the growth rate of the $\ell=1$ mode remains unchanged because the stress trace $t^i_{\,i}$ stays uniform across the sphere for $\ell=1$ perturbations (in the linear regime).
In fact, only higher order modes generate area shear deformation in the linear regime, which allows for local variation of $t^i_{\,i}$ and a corresponding variation of molecular lifetime $1/k_{off}(t^i_{\,i})$ at the cortex (see Eq.~\eqref{eq:StressTracePert} and Supplementary Section~3).

Admitting for two molecular regulators with mixed mechanosensitivity, we found that oscillatory  higher order modes ($\ell>1$) can be generated as transient patterns on the sphere (Fig. \ref{fig:Fig3}). 
For a flat two-dimensional active surface, stable oscillatory solutions are predicted by linear stability analysis for adequate parameter regimes (see Supplementary Section~4).

Motivated by the myosin-rich contractile ring that constricts cells during cell division, we asked to which extent  active spherical surfaces in our model can be constricted by one molecular regulator with catch bond mechanosensitivity.
We observed that amplification of $\ell=2$ mode perturbations in ring conformation can give rise to self-organized central accumulation of molecular regulators and central constriction of the cell. 
We found that this constriction increases with the ratio $\eta_s/\eta_b$ between area shear and bulk viscosity.
However, we further observed that corresponding ring conformations of the cortex are not stable with regards to small asymmetric concentration perturbations in the nonlinear regime; over time, ring conformations transition into other patterns, such as two symmetric concentration maxima at the poles, via asymmetric ring slippage (see Fig. \ref{fig:Fig3}e,f).  

To investigate if `ring' conformations of the active surface could be stabilized in the nonlinear regime, we further tested a modification of the model that exclusively affects the nonlinear range of concentration and shape perturbations. In particular, we investigated the case of area shear viscosities that depend on the local concentration of molecular regulators as motivated by experimental observations \cite{gard04}. 
We found that the `ring' conformation of the active cortex can be stabilised by this effect and even ring centering can be achieved over time (Fig.~\ref{fig:Fig4}d,e). 

Numerous experimental findings have highlighted the key role of actin cross-linkers as key regulators of actin cortex mechanics in cells. In addition, experimental evidence is increasing that actin cross-linker binding dynamics is by itself mechanosensitive \cite{yao13, schi16, luo13, hoss20, mull20}. 
Here, we investigated  the influence of mechanosensitive molecular regulators on pattern formation and deformation of active spherical surfaces.
Our analysis shows that mechanosensitivity of molecular regulators significantly enriches the pattern spectrum of mechanochemical self-organization in the regime of large hydrodynamic length scales. Thus molecular mechanosensitivity of cortical regulators may serve cells as a mechanism  to tune cortical pattern formation adapted to cellular function.

\section*{Acknowledgments}
We thank Frank Jülicher, Christoph Weber, Stephan Grill and Axel Voigt for support of the project and fruitful discussions on the topic.
SA and EFF acknowledge financial support from the DFG in the context of the Forschergruppe FOR3013, projects AL 1705/6-1 (SA) and FI 2260/5-1 (EFF).
E.F.-F. was further supported by the Deutsche Forschungsgemeinschaft under Germany's Excellence Strategy, EXC-2068-390729961, Cluster of Excellence Physics of Life of TU Dresden.
Simulations were performed at the Center for Information Services and High Performance Computing (ZIH) at TU Dresden.



\bibliographystyle{biophysj}
\bibliography{CrossLinker}
\end{document}


\title{Supplementary: \\ 
On the role of mechanosensitive binding dynamics in the pattern formation of active surfaces}
\author{M. Bonati}
\affiliation{Cluster of Excellence Physics of Life, Technische Universit\"at Dresden, Dresden, Germany}
\author{L.D. Wittwer}
\affiliation{Faculty of Mathematics and Informatics, Technische Universtit\"at Bergakademie Freiberg, Freiberg, Germany}
\affiliation{Faculty of Informatics/Mathematics, Hochschule f\"ur Technik und Wirtschaft, Dresden, Germany}
\author{S. Aland}
\email{Corresponding author: sebastian.aland@math.tu-freiberg.de}
\affiliation{Faculty of Mathematics and Informatics, Technische Universtit\"at Bergakademie Freiberg, Freiberg, Germany}
\affiliation{Faculty of Informatics/Mathematics, Hochschule f\"ur Technik und Wirtschaft, Dresden, Germany}
\author{E. Fischer-Friedrich}
\email{Corresponding author: elisabeth.fischer-friedrich@tu-dresden.de}
\affiliation{Cluster of Excellence Physics of Life, Technische Universit\"at Dresden, Dresden, Germany}
\affiliation{Biotechnology Center, Technische Universit\"at Dresden, Dresden, Germany}
\affiliation{Faculty of Physics, Technische Universit\"at Dresden, Dresden, Germany}

\maketitle
\section{Linear stability analysis}
\label{sec:LinStab}
A derivation of the linear stability analysis of the active fluid model on a sphere has been previously presented by Mietke {\it et al.} \cite{miet19b}.
For our study, we restrict this previously presented analysis to the experimentally motivated parameter regime for a cellular cortex.  In turn, we will anticipate 
\begin{itemize}
\item[i)] a negligible external bulk viscosity ($\bar \eta=0$): $\bar \eta$ models the viscosity of the aqueous medium around the cell  while cortical and cytoplasmic viscosities were estimated to be orders of magnitude higher  \cite{fisc16, hoss21, kalw11,dani06}. 

\item[ii)] a negligible cellular surface tension in the absence of molecular regulators ($\gamma=0$, see \cite{miet19b}):  cellular surface tension was shown to be strongly dependent on the activity of molecular motors \cite{fisc14, fisc16, tine09}.

\item[iii)] a negligible contribution of curvature to the active in-plane stress in the surface ($\xi^\prime=0$, see \cite{miet19b}):  cell surface tension was shown to be largely independent of the degree of cell confinement induced by cell squishing with the cantilever of an atomic force microscope in spite of growing curvature of the free-standing cell surface \cite{fisc14}. 

\item[iv)] a negligible bending stiffness ($\kappa=0$): This approximation corresponds to the well-established membrane approximation for thin closed shells \cite{land86}. \end{itemize}

In the following, we discuss the linear stability of the homogeneous stationary state in which the active surface is given by a sphere of radius $R_0$, the concentration of molecular regulators is $c_0$ on the sphere and there are no flows, i.e. $\textbf{v}=0$. \\
Following Mietke {\it et al.}\cite{miet19b}, we express the shape, concentration and flow perturbations as
\begin{eqnarray}
    \delta R= \sum_{\ell ,m} \delta R_{\ell m}Y_{\ell m}, \quad
    \delta c= \sum_{\ell ,m} \delta c_{\ell m}Y_{\ell m}, \quad
     \delta \textbf{v}_\parallel = \sum_{\ell ,m} ( \delta v_{\ell m}^{(1)}\mathbf{\Psi}^{(\ell m)} + \delta v_{\ell m}^{(2)}\mathbf{\Phi}^{(\ell m)} )  \quad.
\end{eqnarray}
where $Y_{\ell m}(\theta, \varphi)$, $\mathbf{\Psi}^{(\ell m)}(\theta, \varphi)$ and $\mathbf{\Phi}^{(\ell m)}(\theta, \varphi)$ are scalar and vector spherical harmonics, respectively.
Including mechanosensitive unbinding of molecular regulators, one finds from the linear stability analysis of dynamic Eqn.~(2)-(5), main text,  the following system of linear equations in the perturbation amplitudes $\delta c_{\ell m}, \dvone, \dvtwo$ and $\delta R_{\ell m}$
\begin{eqnarray}
\frac{\eta_s}{R^2_0}(1-\ell)(\ell+2) \delta v_{\ell m}^{(1)} + \frac{\eta_b}{R^2_0}\left( 2 \delta \dot{R}_{\ell m} - \ell(\ell+1)\delta v_{\ell m}^{(1)}\right) + \frac{\xi}{R_0}f'(c_0)\delta c_{\ell m}  &=&    
\frac{\eta}{R_0} (1+2\ell)\dvone - 3 \frac{\eta}{R_0}  \frac{\drdot}{\ell} \hspace{2cm} \label{eq:LinStab1}\\
\frac{\eta_s}{R^2_0}(1-\ell)(\ell+2)\dvtwo &=& \frac{\eta}{R_0}(\ell-1)\dvtwo \label{eq:LinStab2}\\
\frac{2\eta_b}{R^2_0}(2 \drdot - \ell(\ell+1)\dvone) + \frac{2 \xi}{R_0} f'(c_0)\delta c_{\ell m} + \frac{\xi }{R^2_0}f(c_0)(\ell+2)(\ell-1)\delta R_{\ell m} &=& \label{eq:LinStab3}\\ 
&& \hspace{-3cm} 
\frac{3\eta}{R_0} (\ell+1) \dvone - \frac{\drdot}{\ell(\ell+1)R_0}(\eta(4 + 3\ell + 2\ell^2)\ell + 3\eta) \nonumber \quad.
\end{eqnarray}
Here, Eqn.~\eqref{eq:LinStab1} and \eqref{eq:LinStab2} are derived from the balance of tangential forces in the surface, while Eq.~\eqref{eq:LinStab3} emerges from the balance of forces normal to the surface \cite{miet19b}. Furthermore, the linearized time evolution of the concentration field gives
\begin{eqnarray}
\delta \dot c_{\ell m} + \frac{c_0}{R_0}(2\drdot-\ell(\ell+1)\dvone)+\frac{D}{R_0^2}\ell(\ell+1)\delta c_{\ell m}+k_{off}(\sigma_0)\delta c_{\ell m}+k_{off}'(\sigma_0)\delta t^i_{\,i}\,c_0&=&0 \quad,
\label{eq:LinStab4}
\end{eqnarray}
where $\sigma_0=t^i_{\,i,0}=2\xi f(c_0)$ is the trace of the stress in the steady state ($c=c_0$) and the perturbation of the trace of the in-plane stress $\delta t^i_{\,i}$ is given by 
\begin{equation}
\delta t^i_{\,i}=\frac{2\eta_b}{R_0}\left(2\drdot-\ell(\ell+1)\dvone\right) +2 \xi f'(c_0)\delta c_{\ell m} \quad.  \label{eq:StressTracePert}
\end{equation}
Anticipating that perturbations either grow or shrink exponentially in time with a growth rate $\lambda_{\ell m}$, we can replace time derivatives $\delta \dot{R}_{\ell m}$ and $\delta \dot c_{\ell m}$ in the above Eqn.~\eqref{eq:LinStab1}-\eqref{eq:LinStab4} by $\lambda_{\ell m} \delta {R}_{\ell m}$ and $\lambda_{\ell m} \delta  c_{\ell m}$.
Eq.~\eqref{eq:LinStab2} shows that the mode amplitude $\dvtwo$ has to vanish.
Remaining Eqn.~\eqref{eq:LinStab1} and \eqref{eq:LinStab3}-\eqref{eq:StressTracePert} allow to calculate the value of the growth rate $\lambda_{\ell m}$. It turns out that $\lambda_{\ell m}$ is independent of the mode number $m$. Therefore, we  write $\lambda_{\ell}$ throughout this manuscript instead of $\lambda_{\ell m}$.

For the more general case of several molecular regulators in the model, the above linear stability analysis can be extended in a straight forward manner; in this case, a time evolution equation like Eq.~\eqref{eq:LinStab4} has to be taken into account for all concentration fields in the model. Furthermore, the term $f'(c_0) \delta c_{\ell m}$ in Eqn.~\eqref{eq:LinStab1}, \eqref{eq:LinStab3} and \eqref{eq:StressTracePert} has to be replaced by $\partial_{\bf c} f({\bf c_0}) \cdot\delta {\bf c}_{\ell m}$, where $\bf c$ is the vector of concentration fields in the model and $\delta {\bf c}_{\ell m}$ is the vector of amplitudes of perturbations of the respective concentration field in  harmonic mode ($\ell, m$).

\section{The growth rate in the limit of vanishing bulk viscosity}
\label{sec:GrowthRateLH}
For one molecular regulator, one finds for the growth rate $\lambda_{\ell}$  of a harmonic perturbation with mode number $\ell>0$ (independent of the value of $m$) in the regime of  vanishing bulk viscosity
\begin{eqnarray}
\lambda_{\ell}=\frac{1}{8 R_0^2 \eta_b \eta_s}
\Bigg[ 4 \eta_s \left(c_0 R_0^2 \xi f'(c_0)-\eta_b \left(D l (l+1)+R_0^2 k_{off}^0\right)\right)+R_0^2\xi f(c_0) \left(2 \eta_s-l (l+1) \left(\eta_b+\eta_s\right)\right) & \notag\\
&\hspace{-14cm} 
 \pm \sqrt{\Big\{-8 R_0^2 \xi f(c_0) \eta_s \Big[ c_0 R_0^2 \xi  f'(c_0) \left((l-1) (l+2) \eta_s (4 k_{off}^\prime(\sigma_0) \eta_b+1)-l (l+1) \eta_b\right) \hspace{2cm}}         \notag\\
& \hspace{-10cm}  +\eta_b \left(l (l+1) \left(\eta_b+\eta_s\right)-2 \eta_s\right) \left(D l (l+1)+R_0^2 k_{off}^0\right)\Big]           \notag\\
& \hspace{-14cm}  +16 \eta_s^2 \left(\eta_b \left(D l (l+1)  +R_0^2 k_{off}^0\right)-c_0 R_0^2 \xi  f'(c_0)\right)^2          +R_0^4 \xi^2 f(c_0)^2 \left(l (l+1) \left(\eta_b+\eta_s\right)-2 \eta_s\right)^2		\Big\}	\Bigg]			\quad, \notag   
 \end{eqnarray}
 where $\sigma_0=2\xi f(c_0)$, $k_{off}^0$ and $c_0$ are the trace of the stress, the unbinding rate and the concentration in the steady state, respectively.
For $\ell=0$, we find 
 $$\lambda_{0}= -\frac{3 k_{on}}{R}-k_{off}^0-k_{off}'(\sigma_0)  2 \xi f'(c_0)c_0 \quad.
 $$

\section{The strain rate tensor in the limit of vanishing bulk viscosity}
\label{sec:StrainRateMixed}
For vanishing bulk viscosity (limit of diverging hydrodynamic length scale), the  strain rate tensor $v^i_{\,j}$ reads in spherical coordinates $(\theta, \varphi)$ to first order
\begin{eqnarray}
\frac{\xi   f'\left({c}_0\right) \delta {c}_{\ell m}}{2 \eta_b \left(\xi  f\left({ c}_0\right)+2\lambda_{\ell} \eta_s\right)} 
\left(
\begin{array}{cc}
 \xi  f\left({c}_0\right) Y_{\ell m}^{(2,0)}-2\lambda_{\ell} \eta_s Y_{\ell m} & 0\\
0 & 
\xi  f\left({ c}_0\right) \left(\cot (\theta )  Y_{\ell m}^{(1,0)}+\csc ^2(\theta ) Y_{\ell m}^{(0,2)}\right)-2\lambda_{\ell} \eta_s Y_{\ell m}
\end{array}
\right),
\end{eqnarray}
where we have omitted the argument $(\theta, \varphi)$ of  the spherical harmonic function $Y_{\ell m}(\theta, \varphi)$ for brevity. Correspondingly, the strain rate tensor is proportional to the identity matrix for $\ell=1$, i.e. that the shear contribution to the strain rate vanishes for $\ell=1$ to first order.

\section{Opposite mechanosensitivity of two molecular regulators in flat two-dimensional space}
\label{sec:FlatSurface}
Consider a flat, thin active surface with two molecular regulators of surface contractility described by concentration fields $c_1$ and $c_{2}$. Let $c_{1,free}$ and $c_{2,free}$ denote the concentration fields of freely diffusing unbound motor and passive cross-linker proteins.  Correspondingly, we have the following reaction dynamics 
\begin{eqnarray}
\partial_t c_1 + \nabla\cdot(c_1\bv)= k_{on,1}\cmfree -k_{off,1}(\tr(\bs)) c_1 + D_1 \Delta c_1 \label{eq:MotorConc}\\
\partial_t c_2 + \nabla\cdot(c_2\bv)= k_{on,2}\ccfree -k_{off,2}(\tr(\bs)) c_2 + D_2 \Delta c_2  \label{eq:CLConc} \quad.
\end{eqnarray}
As there is no curvature, the in-plane mechanical stress $\bs$ is  
\begin{eqnarray}
\bs=\eta_s \left(((\nabla \bv)^T+\nabla \bv) -\tr(\nabla \bv) \mathbb{1}\right)+\eta_b \tr(\nabla \bv) \mathbb{1}+\xi f(\boldsymbol{c})  \mathbb{1}.  \quad,
\end{eqnarray}
where  $\boldsymbol{c}=(c_1,  c_2  )$.
Consider a perturbation in the form of a Fourier mode with wave vector $\bk$, i.e.
$
\delta c_{1/2}=\delta c_{1/2}^k \exp(i {\bk} \cdot {\br}),\quad \delta \bv=\delta\bv^k \exp(i {\bk}\cdot {\br})
$
with $k>0$ and $\br=(x,y)$. By the force balance equation  $\nabla\cdot \bs=0$, we have 
\begin{equation}
i \bk\cdot \delta\bs=-\eta_s \bk^2 \bv - \eta_b (\bk\cdot \bv) \bk+ i \bk \xi (\partial_{\boldsymbol c}  f(\boldsymbol c_{0})\cdot\delta \boldsymbol{c})=0.
\end{equation}
Therefore, the velocity contribution orthogonal to $\bk$ has to vanish.
Further 
\begin{equation}
\bk\cdot\bv =\frac{i\xi (\partial_{\boldsymbol c}  f(\boldsymbol c_{0})\cdot\delta \boldsymbol{c})}{(\eta_s+\eta_b)}  \label{eq:ForceBalance}
\end{equation}
and, correspondingly, we have 
\begin{equation}
\tr(\delta\bs)=2\eta_b\, i{\bk}\cdot \bv+2 \xi \partial_{\boldsymbol c}  f(\boldsymbol c_{0})\cdot\delta \boldsymbol{c} =- \frac{2\eta_b\xi (\partial_{\boldsymbol c}  f(c_{0})\cdot\delta \boldsymbol{c})}{(\eta_s+\eta_b)}+ 2 \xi\partial_{\boldsymbol c}  f(\boldsymbol c_{0})\cdot\delta \boldsymbol{c}= \frac{2\eta_s \xi \partial_{\boldsymbol c} f(\boldsymbol c_{0})\cdot\delta \boldsymbol{c}}{(\eta_s+\eta_b)} \label{eq:TrSigma}
\end{equation}
The dynamic equation for the perturbation is
\begin{eqnarray}
\partial_t \delta c_1 + i c_{0,1}\bk\cdot\bv &=&  -c_{0,1}k_{off,1}^\prime(\sigma_0) \tr(\delta \bs) -k_{off,1}(\sigma_0) \delta c_1  - D_1 k^2 \delta c_1  \label{eq:PertDyn1}\\
\partial_t  \delta c_2 +i c_{0,2}\bk\cdot\bv &=& -c_{0,2} k_{off,2}^\prime(\sigma_0)\tr(\delta \bs) -k_{off,2}(\sigma_0) \delta c_1 - D_2 k^2 \delta c_2        \quad, \label{eq:PertDyn2}
\end{eqnarray}
where $\sigma_0=2 \xi f({\bf c}_0)$ is the trace of the stress in steady state. In the following, we will omit the argument $\sigma_0$ of $k_{off,1/2}$ and $k^\prime_{off,1/2}$ for brevity.
Using Eqn.~\ref{eq:ForceBalance} and \ref{eq:TrSigma} in Eqn.~\eqref{eq:PertDyn1} and \eqref{eq:PertDyn2}, we find finally
\begin{eqnarray}
\partial_t  \delta \boldsymbol{c}& = & \notag
\begin{pmatrix}
 \frac{c_{0,1}\xi \partial_{c_1}\! f({\bf c_{0}})(1-2k_{off,1}^\prime\eta_s)}{(\eta_s+\eta_b)}  
 -k_{off,1}  - D_1 k^2 
 & \frac{c_{0,1}\xi \partial_{c_2}\! f({\bf c_{0}})(1-2 k_{off,1}^\prime\eta_s)}{(\eta_s+\eta_b)}  \\
 \frac{c_{0,2}\xi \partial_{c_1}\! f({\bf c_{0}})(1-2 k_{off,2}^\prime\eta_s)}{(\eta_s+\eta_b)}   &  
 \frac{c_{0,2}\xi \partial_{c_2}\! f({\bf c_{0}})(1-2\ k_{off,2}^\prime\eta_s)}{(\eta_s+\eta_b)} -k_{off,2} - D_2 k^2  
 \end{pmatrix}\delta \boldsymbol{c} \quad.
 \end{eqnarray}
Let's anticipate that $D=D_1=D_2$, $\partial_{c_1}\! f({\bf c}_{0})=\partial_{c_2}\! f({\bf c}_{0})$ and $c_{0,1}=c_{0,2}$. In this case, we find 
 \begin{eqnarray}
\partial_t \delta \boldsymbol{c}& = & \notag
\left( \chi  - D k^2\right)\delta \boldsymbol{c} -
\begin{pmatrix}
2\chi\eta_s k_{off,1}^\prime +k_{off,1}\!   &
2\chi\eta_s k_{off,1}^\prime \\
2\chi\eta_s k_{off,2}^\prime &
 2\chi\eta_s k_{off,2}^\prime +k_{off,2} 
 \end{pmatrix}\delta \boldsymbol{c},
\end{eqnarray}
where $\chi= c_{0,1}\xi \partial_{c_1}\! f({\bf c}_{0})/(\eta_s+\eta_b)$.
The corresponding eigenvalues of the time derivative operator are 
 \begin{eqnarray}
 \lambda_{k,1/2}&=& -D k^2+\chi  (1- \eta_s ( k_{off,1}^\prime+ k_{off,2}^\prime))-\frac{1}{2} (\koffo+\kofft) \pm \\
 && \sqrt{\eta_s^2 \chi ^2 ( k_{off,1}^\prime+ k_{off,2}^\prime)^2+  \chi \eta_s  (k_{off,1}^\prime - k_{off,2}^\prime) (\koffo-\kofft)+\frac{1}{4}(\koffo-\kofft)^2} \notag.
 \end{eqnarray}
Let's consider the particular case of a catch bonding molecular regulator described by $c_1$, i.e. $\koffo^\prime<0$ and slip bonding molecular regulator described by $c_2$ with $\kofft^\prime =- \koffo^\prime$.
In this case, 
\begin{eqnarray}
 \lambda_{1/2}&=& -D k^2+\chi -\frac{1}{2} (\koffo+\kofft) \pm  \sqrt{\frac{1}{4}\Delta\koffo( \Delta\koffo-8 \chi \eta_s  | k_{off,1}^\prime |)} \notag\quad,
 \end{eqnarray}
 where $\Delta\koffo=(\koffo-\kofft)$. For the case $\Delta\koffo>0$, i.e. if $\koffo> \kofft$, we obtain an imaginary contribution in the eigenvalue if 
 $$\frac{8 \eta_s c_{0,1}\xi \partial_{c_1} f({\bf c_{0}})}{(\eta_s+\eta_b)} |k_{off,1}^\prime|> \Delta\koffo$$
 and therefore, oscillating solutions are predicted to emerge.

\section{Numerical Method}
\label{sec:NumMethod}
To examine the behaviour of the system away from the linear stability region, we solve the Eqn.~(2)-(4) in an axisymmetric scheme. 
A detailed description of the numerical method can be found in \cite{witt22}. Here, we summarize briefly the main features of the method.

The method is based on a finite-element framework, implemented in the AMDiS toolbox \cite{Vey07, witkowski2015software}. Rewriting the equations in axisymmetric form reduces the effective computational domain to two dimensions, see \cite{mokbel2020ale}. 
The viscous fluid surface is embedded into two bulk fluids (intracellular and extracellular) to simplify the solution procedure. Intracellular bulk fluid viscosity is set to $R_0\eta/\eta_b=0.1$. We also included an extracellular bulk Stokes fluid with viscosity $R_0\bar\eta/\eta_b=0.001$ in the numerical model. The finite viscosity of this external bulk fluid ensured stability of the numerical model, but the value was chosen so small that it did not influence velocity fields on the active surface. By comparison, $\bar\eta$ was set to zero in the theoretical calculations.
The two-dimensional numerical domain is discretized by using a surface grid for the cell surface and a bulk grid for the surrounding fluids. Both grids match at the surface by construction. Grid size is set to 0.08 at the surface and coarser afar.

The time stepping is constructed by operator splitting, i.e. hydrodynamics and surface concentration equations are solved subsequently in every time step. The incompressible Stokes equations in the surrounding fluids are discretized by standard Taylor-Hood elements (polynomial basis functions of degree two and one for velocity and pressure, respectively). The hydrodynamic equations for both fluid phases are assembled in a unified way following \cite{aland2021unified}, leading to a continuous velocity field, but allowing for a discontinuous pressure field across the cell surface. 

The surface viscous stress is treated explicitly and enters the bulk Stokes equations as a boundary condition. The explicit treatment imposes time step restrictions which are handled by underrelaxation (relaxation factor $\omega = 10^{-3}$). 
The concentration equation is solved on the surface grid and discretized with piece-wise linear basis functions. A conservative weak form of the concentration equations is employed to ensure exact mass conservation under strong tangential surface flows.

To accommodate cellular shape changes and lateral translation of the cell, the discretization is performed on a moving finite-element grid using the Arbitrary-Lagrangian-Eulerian (ALE) method\citep{de2021numerical}. Normal movement of the surface grid is imposed conforming with the flow, while tangential grid movement is chosen such that grid distortion is limited. The grid of the surrounding fluids follows surface motion by harmonically extending surface grid movement into the fluid phases. 

A preceding time step study was conducted to ensure time stepping errors are small at the chosen time step size of $\Delta t \in [10^{-7}, 10^{-5}]$ depending on occurring stress magnitude. 
The evolution of patterns was found robust with respect to time stepping errors. Yet, we find that time step size still somewhat influences the time scale at which the instability develops and changes in patterns occur.
Further details about the numerical method can be found in \cite{witt22}.

In our study, simulations were used to verify predictions about the system pattern forming behavior according to the linear stability analysis (Fig.~1, main text). To this end, we were running simulations for different model parameters with initial  (dimensionless) concentrations corresponding to the  steady state concentration $1$ perturbed by an axisymmetric concentration profile with random values drawn from the interval  $\left[-5\cdot 10^{-3}, 5\cdot 10^{-3}\right]$ (Fig.~1, main text). 
Simulations were carried out for a total simulation time of $1$.
To decide about whether the system was evolving as stable or unstable (see red and green dots in Fig. 1a,b, main text), we calculated the difference $\Delta c$ between the maximum and the minimum concentration on the sphere as a function of time.
If at time $\hat t=1$ the slope of $\Delta c$ was positive, the simulation was classified as unstable, otherwise stable.

\begin{figure}[h]
\centering
\includegraphics[width=14cm]{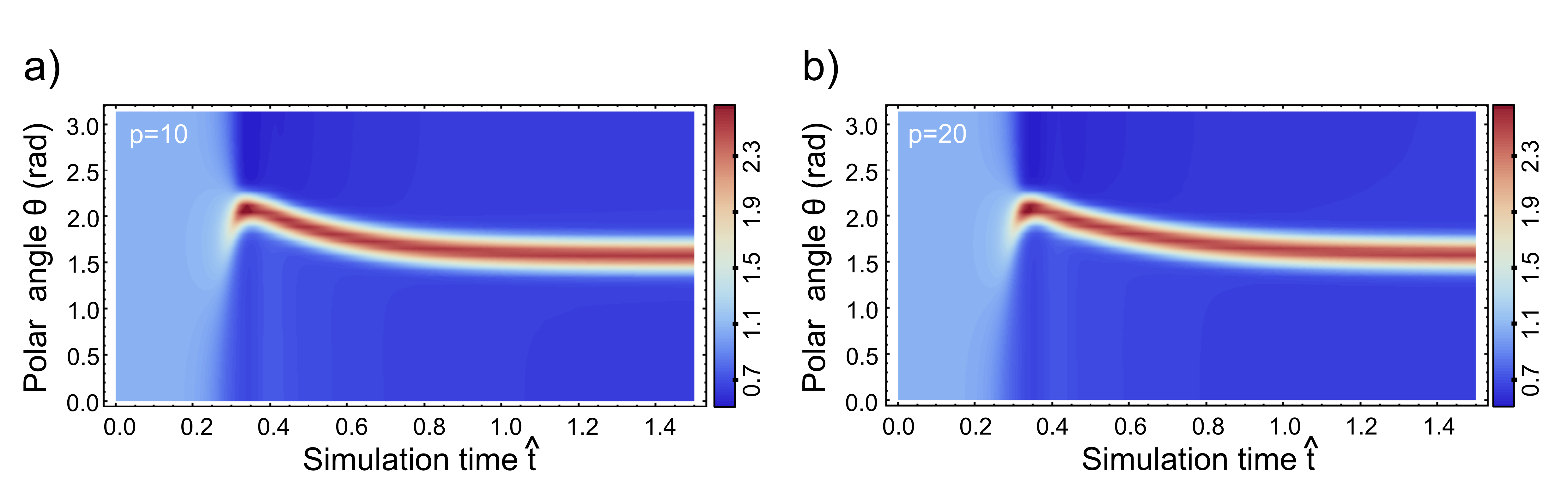}
\caption{
\textbf{Concentration-dependent shear viscosities consolidate ring formation in model cells with catch bonding molecular regulators.}
Kymographs showing the concentration evolution in the presence of strong concentration-dependent viscosity  $\eta_s=\eta_s^0 (p+1)c^2/(p c_0^2+c^2)$ with $p=10$ (panel a) and $p=20$ (panel b).  The initial concentration was chosen as the steady state concentration $c_0$ perturbed with a small axisymmetric concentration profile with random values in the interval $\left[-5\cdot 10^{-3}, 5\cdot 10^{-3}\right]$. 
Dimensionless parameters used were, 
$\eta_s^0=0.25$, $k_{on}R_0/D=200$, $\Delta \hat t = 2.5\cdot 10^{-6}$,
$R_0\eta/\eta_b=0.1$, 
$\sigma_c R_0^2/(\eta_b D)=100$, $\xi R_0^2/(\eta_b D)=125$ and $k_{off}^0 R_0^2/D=100$ respectively. 
\label{fig:FigS1}}
\end{figure}

\newpage
\bibliographystyle{biophysj}
\bibliography{CrossLinkerSupp}